\documentclass[pra, onecolumn, showpacs,aps]{revtex4}
\usepackage[dvips]{graphicx}
\usepackage{bm}

\newcommand{\xcaret}{\bm{\hat{x}}}
\newcommand{\ycaret}{\bm{\hat{y}}}
\newcommand{\zcaret}{\bm{\hat{z}}}
\newcommand{\ucaret}{\bm{\hat{u}}}

\begin{document}

\title{Orientation control of rodlike objects by flow}

\author{C. Tannous}
\affiliation{Laboratoire de Magn\'etisme de Bretagne CNRS-FRE 3117, \\
Universit\'e de Bretagne Occidentale, 29285 Brest, France.}

\begin{abstract}
Suspensions of rodlike objects in a liquid 
are encountered in many areas of science and technology and 
the need to orientate them is extremely important to enhance or inhibit 
certain chemical reactions between them, 
other chemicals or with the walls of vessels holding the flowing suspension.
Orientation control is feasible by altering
velocity and nature of the liquid, its flow and the geometry of the channel containing the flow.
In this work we consider the simplest possibility of orientation control with flow in two dimensions
on the basis of the orientation distribution of rodlike objects. The simple differential
equation satisfied by the orientation probability density function is derived and its solution discussed. 
In addition, we show how birefringence and dichroism ("Maxwell effect") of the suspension provide a direct experimental test of orientation control. \\

\vspace{0.5cm}


\end{abstract}


\maketitle

\section{Introduction}
In many physical, chemical, biological processes, the behavior and orientation of 
rodlike objects (such as fibers, nanotubes, nanowires, DNA, macromolecules) 
suspended in a flowing liquid affect the transport, rheology, chemical and hydrodynamic 
characteristics of the suspension. 
Orientation control for the sake of aligning or separating the
rodlike objects (RLO), enhancing 
reaction between  them, with other chemicals or vessel walls is important in many areas of science and technology
such as sedimentation, blood flow, pulp and paper, polymer processing, liquid crystal flow, microfluidic devices, ferrofluids...

This work is relevant to students who have completed an undergraduate Statistical Physics course 
of the Reif~\cite{Reif} level and are interested in applications of stochastic processes 
or graduate students whose level corresponds to Landau and Lifshitz course~\cite{Landau} and are interested in the general applications of Fluid Mechanics. 
The history of dilute solutions of RLO (from the polymeric point of view) is detailed in
Chapter 8 of Doi and Edwards book "The Theory of Polymer Dynamics" \cite{Doi} as well as
in a review paper by Forest et al. \cite{Forest}. 
Students interested in various other materials and applications are encouraged to look at the 
{\it "Soft Matter"} four volume series edited by G. Gompper and M. Schick~\cite{soft}.

In this work, we are interested in dilute suspensions of rigid and long RLO with a  negligible 
cross-sectional area and concentrate on the study of orientation
control by flow in two dimensions. Our goal is to investigate in 
the simplest, yet physical case, and with a self-contained pedagogical point of view,
the effect of liquid shear flow on the rotational diffusion of a dilute suspension of 
non-interacting RLO in a planar contraction. We derive in a straightforward manner
the main equation describing the statistical distribution of orientation in two dimensions (2D), 
solve it analytically and show how orientation control might be detected optically with 
a measurement such as birefringence~\cite{Hecht} (double refraction) and dichroism of the suspension.

Objects suspended in flow undergo two types of motion: smooth motion due to the average fluid velocity field and
erratic random motion produced by the fluctuating fluid velocity, temperature and inertia driven motion. 
The resulting change in the suspension 
microstructure can have a significant effect on the mechanical, thermal, optical, electrical,
magnetic and chemical properties. The equation that accurately models the orientation state of non-interacting
RLO in hydrodynamic nonhomogenous flow  depends on the rotational Peclet number $\alpha$
that represents the interplay between the randomizing effect of temperature induced rotational 
diffusion and the orientating effect of streamwise mean rate of strain due to the liquid flow.

\begin{figure}[htbp]
\centering
\scalebox{1.2}{\includegraphics[angle=0]{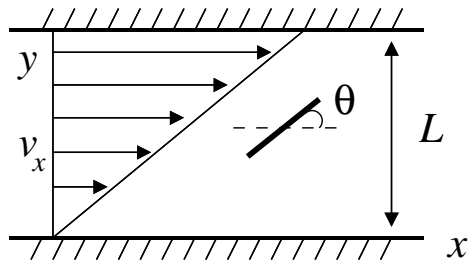}}
\caption{2D geometry of the shear flow and the RLO of length $\ell$ making an angle $\theta$ with 
the flow direction along $x$. The shear rate of the flow is $\dot{\gamma}=\frac{\partial v_x}{\partial y}$.
The fluid velocity profile $v_x$ is shown explicitly as a function of the coordinate $y$. 
In the frame moving with the rod, the velocity field is antisymmetric (resp. symmetric) above 
(and resp. below) the rod.} 
\label{shear}
\end{figure}

Two types of flow are of particular importance: simple shear flow and elongational 
(extensional) flow. Simple shear flow corresponds to a velocity profile where the gradient in the fluid
flow velocity is constant. In the case of elongational flow the sample is compressed in one 
direction and elongated in the other.  In this work we consider a dilute concentration of 
non-interacting RLO suspended in a liquid performing shear flow exclusively 
in a planar contraction of width $L$ (see fig.~\ref{shear}).

The rotational Peclet number $\alpha$ measuring the relative strength of hydrodynamic interactions and Brownian 
forces is defined mathematically as the ratio of hydrodynamic shear flow and the rotational diffusion constant:

\begin{equation}
\alpha= \frac{\dot{\gamma}}{D_{r}} 
\end{equation}

$\dot{\gamma}$ is the (hydrodynamic) shear rate and $D_{r}$  the (thermal) rotational diffusion coefficient
(see the Appendix for an explanation about the difference between ordinary diffusion~\cite{Reif} and rotational
diffusion).

The flow shear rate defined by $\dot{\gamma}=\frac{\partial v_x}{\partial y}$ 
with $v_x$ the velocity field in the $x$ direction of the flow (see fig.~\ref{shear}) 
is related to translational degrees of freedom whereas $D_{r}$, the Brownian diffusion coefficient, 
governs the rotational motion of the RLO  around its center of mass and hence relates to rotational
degrees of freedom (see Appendix).

The RLO suspended in the liquid are under the action of two orientating
forces: one that comes from the hydrodynamic flow and the other due to thermal effects. 
Despite the apparent simplicity of the problem, its physics is extremely rich and one might stumble upon some unusually complex behaviour; for instance the following set of states may be observed:
\begin{enumerate}
\item An {\it aligned} state where the RLO is aligned with the flow at a fixed angle;
\item A {\it tumbling} state, in which the RLO lies in the shear
($x-y$) plane and rotates about the $z$ axis. 
\item  A {\it wagging} state, in which the RLO lies in the shear
plane, but oscillates between two values.  
\item  A {\it Kayak-tumbling} state, equivalent to the tumbling state,
but in which the RLO is out of the shear plane.
\item  A {\it Kayak-wagging} state where the
RLO is out of plane, but the projection of the RLO on the
shear plane oscillates between two values.
\item  A {\it mixed} state, typically found close to the boundaries
between wagging and tumbling states and where the RLO
undergoes both oscillation and complete rotation. 
\item  A {\it Chaotic} state, in which the RLO angle with the flow
behaves with time in an apparently random way. 
\end{enumerate}

While the above describe the time dependence of the RLO orientation, it would be rather interesting to
evaluate the statistical distribution of the angles the RLO makes with the flow. 
Specializing to 2D,  Boeder~\cite{Boeder} is the first to have studied the RLO stationary angular
distribution problem in a flowing liquid from a theoretical and experimental point of view. 
In the shear flow case, he derived an ordinary 
differential equation (ODE) satisfied by  
$P(\theta)$, the stationary PDF (see Appendix) describing the 
average orientations of the RLO assuming that the motion of RLO, of negligible 
cross-sectional area occurs in the plane of the flow, without any boundary conditions. 

The resulting ODE reads:
\begin{equation}
\frac{d^2}{d\theta^2} P(\theta) + \frac{d}{d\theta} [\alpha \sin^{2}(\theta) P(\theta)] =0
\label{boeder}
\end{equation}

In order to derive the above ODE, the RLO angular time evolution induced by shear 
flow is needed. We assume that the effects of gravity and inertia on the RLO are 
negligible as well as any disturbance motion resulting from the RLO presence.
Taking a unit vector $\bm{\ucaret}$ along the RLO axis, 
$\bm{\ucaret}= \bm{\xcaret} \cos \theta  + \bm{\ycaret} \sin \theta $, its 
time derivative is given by $ \frac{d \bm{\ucaret}}{dt}= \frac{d\theta}{dt} (-\bm{\xcaret} \sin \theta  + \bm{\ycaret} \cos \theta )$. This yields the angular time evolution of the RLO as 
${\zcaret} \frac{d\theta}{dt} ={\bm{\ucaret}} \times  \frac{{d \bm{\ucaret}}}{dt}$
where $\bm{\zcaret}$ is the unit vector perpendicular to the $x-y$ plane. 
In addition, the fluid velocity field  $\bm{v}=v_x \bm{\xcaret} $ is along 
the $x$ direction and depends solely on the coordinate $y$. 
The above simplifying assumptions imply that the
RLO follows the surrounding fluid velocity gradient and that
the time derivative of $\bm{\ucaret}$ along the $x$ direction (see fig.~\ref{shear})
is proportional to the shear rate ($v_x$ change along $y$) and the $u_y$ component (see additional note~\cite{flow}). This means $\frac{d u_x}{dt}=  \dot{\gamma} u_y $, i.e.
$\frac{d \bm{\ucaret}}{dt}=  \dot{\gamma} \sin\theta  \bm{\xcaret} $. 
Therefore, the RLO angular time evolution induced by shear flow becomes:

\begin{equation}
\frac{d\theta}{dt}= -\dot{\gamma}\sin^{2}\theta
\label{angspeed}
\end{equation}

Thermal disturbances to the RLO motion, are accounted for by adding to the above
 a random term
proportional to a time-dependent White noise amplitude $\xi(t)$ 
possessing the following statistical properties: 

\begin{equation}
<\xi(t)>=0 \mbox{  and  } <\xi(t)\xi(t')> = \delta(t-t') 
\end{equation}

where $<\xi>$ is the statistical average of $\xi$, meaning that $\xi(t)$ has zero mean and zero
correlation with another amplitude occuring at a different time.
The angular time evolution becomes:

\begin{equation}
d\theta= A dt + \sqrt{B} \xi(t) dt
\end{equation}

with $A=-\dot{\gamma}\sin^{2}(\theta)$ and $B$ a constant independent of time and the angle $\theta$.
This implies that the deterministic hydrodynamic forces tend to act on 
the RLO rotating it in the shear flow with an average angular speed $\frac{d\theta}{dt}$
while the fluctuating term $\xi(t)$ originating from thermal effects produce random disorientations.

The differential equation $A dt + \sqrt{B} \xi(t) dt$ in the variable $\theta$ containing the
random term $\xi(t) dt$ is called a Langevin equation~\cite{Reif} that can easily be transformed
into a 1D Fokker-Planck equation (see the derivation in Chapter 15 of Reif's book~\cite{Reif} or the paper titled "Simplified Derivation of the Fokker-Planck equation" by 
Siegman~\cite{Siegman} in this Journal):

\begin{equation}
-\frac{d }{d\theta}[A P(\theta)] + \frac{1}{2} \frac{d^2 }{d\theta^2}[B P(\theta)] =0 
\end{equation}

Straightforward comparison with the ODE given by eq.~\ref{boeder} leads to $B= 2 D_{r}$ and in the
limit of no shear ($\alpha=0$) we recover the stationary planar rotational diffusion equation
$D_{r} \frac{d^2}{d\theta^2}  P(\theta)=0$ (see Appendix).

Having derived the Boeder ODE from simple assumptions, this work is meant to provide 
a solution in closed form for the ODE (eq.~\ref{boeder}) using standard analytical 
methods (pertaining to Freshman calculus level) to obtain $P(\theta)$ for any  
Peclet number $\alpha$  and assess later on the impact
with regard to orientation control from the measurement of optical properties.
  
This paper is organised as follows: In Section II, we present an exact analysis of the
ODE (see eq.~\ref{boeder}) to obtain the PDF, for a wide range of $\alpha$ $\in [10^{-4}-10^{8}]$. 
In section III optical properties are evaluated to test liquid flow effect on orientation control. 
Conclusions are given in Section IV.  The Appendix explains in detail the difference between ordinary and 
rotational diffusion.

\section{Analytical solution}
In this section, an analytic procedure for deriving the solution of the
ODE (see eq.~\ref{boeder}) as the probability density function (PDF), $P(\theta)$ is presented.
The PDF solution of the ODE is $\pi$-periodic since the RLO's
are indistinguishable when orientated at $\theta \mbox{ or } \theta+\pi$.
Since the PDF is $\pi$-periodic one might write, in principle, a  Fourier series solution valid for small and large values of $\alpha$. 

Since we are interested in finding an exact solution we differ from this approximate approach by viewing the problem as a boundary condition one, more precisely:

\begin{equation}
P(0)=P(\pi)   \mbox{  and   }  P'(0)= P'(\pi)  
\label{BVP}
\end{equation}

where  $P'(\theta)=\frac{d P(\theta)}{d\theta}$.

Since the PDF ought to be normalised over the interval $[0,\pi]$:

\begin{equation}
\int^{\pi}_{0}{P(\theta)d\theta}=1
\label{norm}
\end{equation}

Consequently, the determination of the PDF is a constrained 
periodic boundary value problem eq.~\ref{boeder}, the constraint originating from
the normalization condition. 

In order to get closed form solutions, we write eq.~\ref{boeder} as  a first-order ODE: 

\begin{equation}
P'(\theta)+ \alpha \sin^{2}(\theta)P(\theta) =C
\label{IVP1}
\end{equation}

with initial condition: $P(\theta=0)=P(0)$.
This is mathematically sound provided the initial value $P(0)$  and the constant $C$ are
known for any  value of $\alpha$.

The constant $C$ appears in eq.~\ref{IVP1} and is equal to $P'(0)$ by substituting 
$\theta=0$ and assuming finiteness of $P(\theta \rightarrow 0 )$ in eq.~\ref{IVP1}).

The formal solution of eq.~\ref{IVP1} may be given generally as:

\begin{equation}
P(\theta) = C \exp[\frac{\alpha}{2} (\frac{\sin(2\theta)}{2}-\theta)]   
 \int_{-\infty}^{\theta} \exp[-\frac{\alpha}{2}(\frac{\sin(2x)}{2}-x)]dx
\label{unstable}
\end{equation}

The lower limit  $-\infty$ is not obvious since the problem is defined over the angular interval $[0,\pi]$.
Analytically, the lower limit  $-\infty$ is the only possibility compatible with the boundary conditions 
given by eq.~\ref{BVP} since the factor $\exp(\alpha x/2)$ appearing in the integrand as given by eq.~\ref{unstable} enlarges the angular interval from $[0,\pi]$ to the interval $]-\infty, \theta]$. The reason behind
this transformation comes from imposing a zero integration constant that can be obtained only when
$x \rightarrow -\infty$ given the factor $\exp(\alpha x/2)$ and the positive sign of $\alpha$.

Performing a change of variables, the solution may then be written as:

\begin{equation}
P(\theta) = \left(\frac{2C}{\alpha}\right) \hspace{1mm} \exp[\frac{\alpha}{4} \hspace{1mm} \sin(2\theta)]  
\int_{0}^{\infty} \exp(-x) \exp[ \hspace{1mm}\frac{\alpha}{4} \hspace{1mm} \sin(\frac{4x}{\alpha}-2\theta)]dx
\label{stable}
\end{equation}
 
The form in eq.~\ref{stable} is superior to the previous form given by eq.~\ref{unstable}, since the 
the exponential factors $\exp(\theta)$ and $\exp(-\theta)$ terms (that can vary rapidly by 
several orders of magnitude) are not used~\cite{Recipes}. 
The constant $C$ is determined from the normalisation condition of the PDF (eq.~\ref{norm}) yielding:

\begin{equation}
\frac{1}{C}= \frac{2}{\alpha}
 \int_{-\frac{\pi}{2}}^{\frac{\pi}{2}} d\theta \int_{0}^{\infty} dx \hspace{1mm} \exp(-x)  
\hspace{1mm} \exp[(\frac{\alpha}{2}) \sin(\frac{2x}{\alpha}) \cos(2\theta -\frac{2x}{\alpha} )]
\label{eq:C}
\end{equation}

From the general expression eq.~\ref{stable}, the initial value $P(0)$ is given by:

\begin{equation}
P(0) = \left(\frac{2C}{\alpha}\right) \int_{0}^{\infty} \exp(-x) \hspace{1mm} \exp[\hspace{1mm} 
\frac{\alpha}{4} \hspace{1mm} \sin(\frac{4x}{\alpha})]dx
\label{eq:P}
\end{equation}

The PDF depends on $\alpha$ which we may want to vary over several orders of magnitude. 

\begin{figure}[htbp]
\centering
\scalebox{0.3}{\includegraphics[angle=-90]{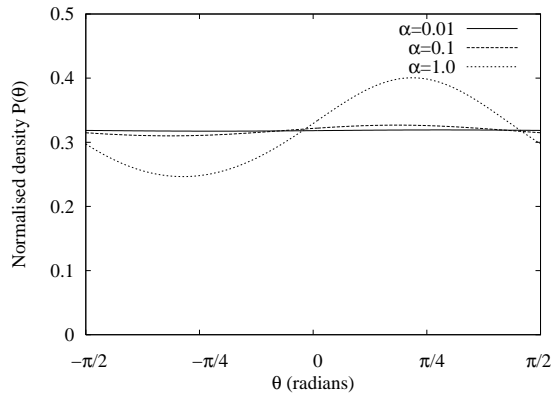}}
\caption{Analytical PDF as a function of $\theta$ for small $\alpha$=0.01, 0.1. and 1.0 normalised
over the interval $[-\pi/2,\pi/2]$.  The isotropic (flat) angular distribution whose value 
is $\frac{1}{\pi}$ over the interval $[-\pi/2,\pi/2]$ is reached as $\alpha$ decreases toward zero.
On the other hand, when $\alpha$ increases, a wavy structure develops about the flat
distribution with a maximum appearing around $\pi/4$. }
\label{small}
\end{figure}

Surprisingly, the difficulty in solving the ODE stems from the fact that its nature may be modified 
when $ \alpha$ increases, with the constraints of $\pi-$periodicity and PDF normalization
maintained. For small values of $\alpha$ the PDF is flat as observed in fig.~\ref{small} and this
can be shown with a simple argument. 

When $\alpha$ approaches zero, we should satisfy the pure rotational
diffusion equation in the plane given by $D_{r} \frac{d^2}{d\theta^2}  P(\theta)=0$ whose solution is
$P(\theta)= a \theta +b $ with $a, b$ constant and determined from $\pi-$periodicity eq.~\ref{BVP} and normalisation condition eq.\ref{norm} yielding $a=0, b=\frac{1}{\pi}$. 

\begin{figure}[htbp]
\centering
\scalebox{0.3}{\includegraphics[angle=-90]{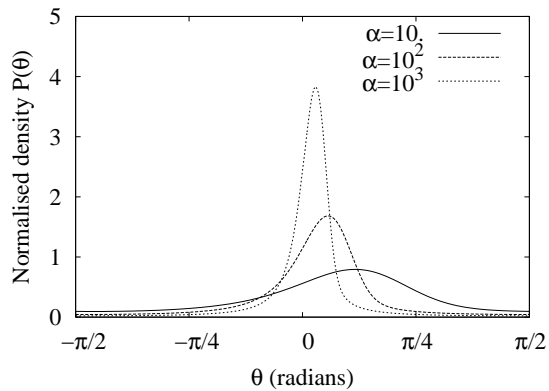}}
\caption{Analytical PDF as a function of $\theta$ for large $\alpha$=10.0, 100. and 1000. normalised
over the interval $[-\pi/2,\pi/2]$. As $\alpha$ increases, the PDF becomes sharply peaked around an angle $\theta_{max}$ that approaches zero.} \label{large}
\end{figure}

In the opposite
case of large  $\alpha$ (as in fig.~\ref{large}) the PDF displays a strong peak around an
angle $\theta_{max}$ indicating that orientation of the RLO is setting in.

\section{Optical properties}
In general, even if a solution of RLO's is optically isotropic at equilibrium, under 
flow conditions, the distribution of orientation will be affected in a way such that
the optical properties become anisotropic because of the existence of some privileged
direction due to the flow. This phenomenon is called flow birefringence or
{\it Maxwell effect}~\cite{Doi}. 

Shear-induced birefringence and dichroism measurements of suspensions of RLO
consists of detecting phase and intensity variation (due to the flow) of light 
travelling across the sample. 

The flow produces a modification in the real and imaginary
part of the optical index $\Delta n', \Delta n"$ of the suspension. 
$\Delta n'$ is extracted from the phase difference between
the transmitted and reference signal whereas $\Delta n"$ is extracted from the attenuation of
light intensity. 
$\Delta n', \Delta n"$ are both proportional~\cite{Doi} to a single quantity, 
$S$ the nematic order parameter~\cite{Straley} encountered in the physics of liquid crystals.

$S$ is 0 for the unorientated (disordered) state and 1 for the fully orientated
(ordered) state. Writing $S=a < \cos^2 \theta > + b $ (with $a,b$ constants to be determined) 
and noting that in the ordered state $< \cos^2 \theta >=1$ since $\theta \sim 0$ whereas in
the disordered state $< \cos^2 \theta >=1/2$.
Hence we find $S=2 < \cos^2 \theta > -1= < \cos 2 \theta > $~\cite{dimension}.

\begin{figure}[htbp]
\centering
\scalebox{0.3}{\includegraphics[angle=-90]{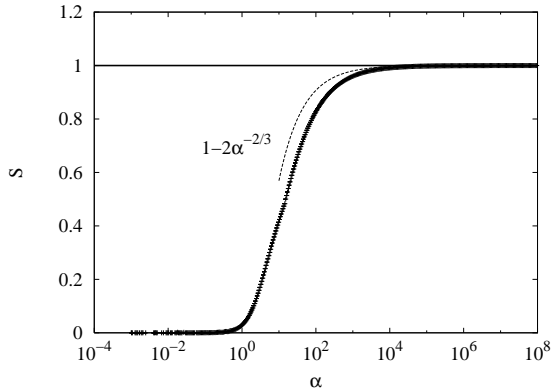}}
\caption{Nematic order parameter $S= < \cos(2 \theta ) > $ versus Peclet number $\alpha$ spanning
the interval $[10^{-4}-10^{8}]$. For small values of  $\alpha$ we have a "disordered" state characterized
by $S \sim 0$ whereas $S$ approaches 1 ("orientated" state) for large values of $\alpha$. The asymptotic
regime  $1- 2 \alpha^{-\frac{2}{3}}$ is shown for comparison.} \label{dichro}
\end{figure}

When $S=< \cos 2 \theta >$  saturates to one, perfect alignment is reached i.e. complete
orientation control is achieved as observed in  Fig.~\ref{dichro} for very large values
of the Peclet number $\alpha$.

When $S$ saturates to 1, the orientational PDF becomes sharply peaked around $\theta_{max}$
the value of the angle that corresponds to the PDF maximum as the Peclet number increases. 
At the value $\theta=\theta_{max}$, the first-order ODE (eq.~\ref{IVP1}) writes:

\begin{equation}
\alpha \hspace{1mm} \sin^{2}(\theta_{max})P(\theta_{max})=C
\label{max}
\end{equation}

Since, it is required that the PDF should be normalised according to eq.~\ref{norm} 
when $\alpha$ is large, we get after using eq.~\ref{max}:

\begin{equation}
 \theta_{max}P(\theta_{max}) \sim \mbox{constant}
\label{area}
\end{equation}

The above is the area under the PDF curve since the distribution function becoming sharply 
peaked around $\theta_{max}$ when $\alpha$ is large, leading to a PDF approximately 
triangular in shape with a height $P(\theta_{max})$  and  a base equal to $2\theta_{max}$. 
This implies that $\theta_{max}$ is close to the standard deviation of the PDF $P(\theta)$
when $\alpha$ is large.

Using eq.~\ref{max} and eq.~\ref{area}, we find the following leading behaviour:

\begin{equation}
\theta_{max}  \sim  \alpha^{-1/3},  \hspace{2mm} P(\theta_{max}) \sim \alpha^{1/3} 
 \end{equation}

Hence $S$ behaves for large $\alpha$ as:

\begin{equation}
S = \int_{-\frac{\pi}{2}}^{\frac{\pi}{2}} P(\theta)    \cos 2 \theta d\theta  
  \sim   \int_{-\theta_{max}}^{\theta_{max}}  P(\theta)   \cos 2 \theta d\theta 
\end{equation}

which yields:

\begin{equation}
S \sim    \theta_{max}  P(\theta_{max}) \cos 2 \theta_{max}
\end{equation}

implying that $S$ behaves as $\cos (2 \theta_{max}) \sim 2 \cos^2 ( \theta_{max}) -1 \sim 1- 2 \alpha^{-\frac{2}{3}}$ 
which is close to the asymptotic behaviour found in fig.~\ref{dichro} and obtained from full numerical integration.

Since $\theta_{max}$ is a measure of the standard deviation (for large values of $\alpha$) in the
fluctuations of the angle $\theta$ about the main flow direction, the RLO orientation controllability
with flow can be analysed with the variation of the angle  $\theta_{max}$ with the Peclet number $\alpha$.

\begin{figure}[htbp]
\centering
\scalebox{0.3}{\includegraphics[angle=-90]{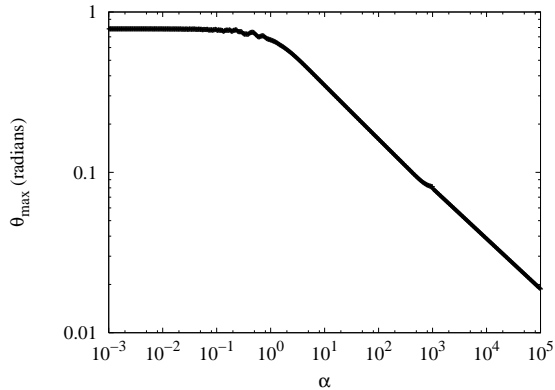}}
\caption{Variation of the angle $\theta_{max}$ (in radians) with the Peclet number $\alpha$. For small values of $\alpha$,  
 $\theta_{max}$ saturates at a value of about 0.8 radian that is close to $\frac{\pi}{4}$ as seen in fig.~\ref{small}. When
$\alpha$ is increased beyond 400, $\theta_{max}$ decreases as $\alpha^{-1/3}$ 
dropping to small values $\le 0.1$ radian indicating that orientation control has started to set in.}
 \label{control}
\end{figure}

In fig.~\ref{control} the variation of  $\theta_{max}$ with the Peclet number $\alpha$
is depicted. From the figure, one infers that orientation controllability is achieved when $\theta_{max}$
drops below a small specified value. For instance, one has to have $\alpha \sim 400$ in order to
achieve $\theta_{max} \le 0.1 $ radian (see fig.~\ref{control}).

\section{Conclusions}
This work is the simplest illustration of the interplay between shear (due to liquid flow)
and rotational diffusion (due to thermal agitation) on the planar orientation of 
a dilute concentration of rigid non-interacting thin and long RLO suspended in a liquid shear flow 
in a simple geometry.

The PDF describing the RLO orientations is analytically  
determined in 2D and accurately evaluated as a function of $\alpha$, 
the rotational Peclet number given by the ratio of shear rate and rotational diffusion constant.

The solutions found for the PDF in the bulk of the flowing liquid for arbitrary values of 
$\alpha$ confirm the validity of the approach used for small values of 
$\alpha$. In this regime flat angular distributions (PDF) are expected: this means we 
have isotropic PDF due to rotational diffusion dominated motion.

On the other, when $\alpha$ increases, we expect sharply peaked angular distributions (PDF)
the peak occuring around an orientation angle $\theta_{max}$: this means we are
in the shear dominated motion case and the PDF is anisotropic with a single mode.

The impact of $\alpha$ on orientation control and its signature through optical properties
is examined and the variation of the nematic order parameter $S$ is obtained for a wide
range  of values of $\alpha$ spanning the interval $[10^{-4}-10^{8}]$.

Since Boeder's pioneering work, many improvements have been made
to remove restrictions on the cross-sectional area and size of the RLO, 
introduce internal vibrational and rotational degrees of freedom within the RLO~\cite{Doi}, 
consider rotational diffusion in 3D, turbulent flow~\cite{Turbulence}, 
more complex flow geometries~\cite{Doi} and interactions between the RLO's.

The Boeder equation provides orientation control requiring a wide variation of the Peclet number
whereas practically a much smaller range is demanded. From fig.~\ref{control} the transition
from $\theta_{max} \sim \pi/4$ (almost flat and isotropic behaviour leading to no alignment) 
to the case $\theta_{max} \sim 0$ (peaked 
behaviour, anisotropic meaning controlled orientation) spans an interval $\alpha$ of $[0.1- 400.]$ in order to switch from no alignment to orientation control in the 0.1 radian range. 
If, on the other hand, we require  $\theta_{max} \le 0.01 $ radian, then $\alpha \sim 6\times 10^{5}$ which
is a very large value.

Practically speaking, a small interval of variation of $\alpha$ is required in order to achieve a fast 
and precise control of alignment (with a small standard deviation), hence the question: \\
Would it be possible, within a model similar to this one and under certain conditions 
to vary slightly $\alpha$ (by a few units only) and achieve a standard deviation smaller than  
0.01 radian? \\ 
Another question is: Would it be possible, within a model similar to this one and 
under certain conditions to vary $\alpha$ and obtain several
peaks (multi-modal) behaviour with several orientations, in other words make a transition
with $\alpha$  from single-mode to multi-mode behaviour?  

\section{{\textbf APPENDIX}: Rotational Diffusion}

Ordinary diffusion describing the random walk of a particle in space is given by: 
\begin{equation}
\frac{\partial f(\vec{r},t)}{\partial t}= D \Delta f(\vec{r},t)
\end{equation}

with $f(\vec{r},t)$ the probability function for the particle to be at position $\vec{r}$ at
time $t$ and $\Delta f$ the full Laplacian of $f$. $D$ is the ordinary (translational) diffusion coefficient. In rotational diffusion, a particle wanders on the surface of the unit sphere such that the diffusion equation  becomes:

\begin{equation}
\frac{\partial f(\theta,t)}{\partial t}= D_{r} \Delta_{\theta} f(\theta,t)
\label{anglap}
\end{equation}

The angular Laplacian $\Delta_{\theta}$  and the probability function  $f=f(\theta,t)$ depends now solely on the angular coordinates represented collectively by $\theta$. $\Delta_{\theta}$ contains (second-order) derivatives with respect to $\theta$ with the radial variable fixed ($r=1$) and 
with the rotational diffusion coefficient $D_{r}$ replacing the ordinary diffusion coefficient $D$.

In the 2D case, the rotational diffusion equation on the unit circle reads:
\begin{equation}
\frac{\partial f}{\partial t}= D_{r} \frac{\partial^2 f}{\partial \theta^2}
\label{diff2D}
\end{equation}

The stationary PDF $P(\theta)$ studied in this work is obtained from the long-time limit:
 $P(\theta)= \lim_{t \rightarrow \infty} f(\theta,t)$.

In contrast to translational diffusion, $D_{r}$ units are not Length$^2$/Time but simply 1/Time since the random walks on the unit sphere are purely angular 
with no dependence on any characteristic length. Similarly to translational diffusion where the average displacement squared $< {[\vec{r}(t)- \vec{r}(0)]}^2 > \sim Dt$ at time $t$, 
in rotational diffusion, we have $< {[\vec{u}(t)- \vec{u}(0)]}^2 > \sim D_{r} t$
with $\vec{u}(t)$ representing the  angular position $\theta(t)$  at time $t$ (while radial 
position $r=1$ is kept fixed for all times).


\begin{thebibliography}{99}
\bibitem{flow} The RLO being rigid, its rotation motion is described by:
$\frac{d \bm{\ucaret}}{dt}= \bm{\Omega} \times \bm{\ucaret}$ where $\bm{\Omega}$ is the rotation vector about
the center. Assuming~\cite{Doi} the RLO is thin, long and follows
the macroscopic velocity gradient, we have in the shear flow case: $\bm{\Omega}= \bm{\ucaret}\times (\bm{\kappa} \cdot \bm{\ucaret}) $
where $\bm{\kappa}$ is the fluid  velocity  gradient tensor with components given by: 
$ \bm{\kappa}_{ij}= \frac{\partial v_i}{\partial x_j}; i,j \in [x,y]$. 
The double cross product yields 
$\frac{d \bm{\ucaret}}{dt}=\bm{\kappa} \cdot \bm{\ucaret}-(\bm{\ucaret \cdot  \kappa \cdot \ucaret)\ucaret}$ 
and the angular time evolution becomes:
${\zcaret} \frac{d\theta}{dt} ={\bm{\ucaret}} \times  \frac{{d \bm{\ucaret}}}{dt}= {\bm{\ucaret}} \times 
(\bm{\kappa} \cdot \bm{\ucaret})$. Note that the tensor $\bm{\kappa}$ has a single component
$ \bm{\kappa}_{xy}=\frac{\partial v_x}{\partial y}= \dot{\gamma}$, the hydrodynamic shear rate, given
the fluid velocity field  $\bm{v}=v_x \bm{\xcaret} $ is along 
the $x$ direction and depends on the coordinate $y$, only (see fig.~\ref{shear}).
\bibitem{dimension}In $n$ dimensions, we have in the disordered state $< \cos^2 \theta >=1/n$ 
since we have $n$ equivalent vector components. Writing as previously $S=a < \cos^2 \theta > + b $ 
and since in the ordered state $< \cos^2 \theta >=1$ 
we find $S=\frac{1}{n-1}(n < \cos^2 \theta > -1)$.
\end{thebibliography}
\end{document}